# PREDATION BY *Atractosteus tropicus* ON *Bufo marinus* AND ITS POSSIBLE AFFECT ON POPULATION IN A LOWLAND TROPICAL WET RAINFOREST

Todd Lewis[1*], Colin Ryall[2], Thomas C. Laduke[3] and Paul B.C. Grant[4]

1 Westfield, 4 Worgret Road, Wareham, Dorset, BH20 4PJ, UK.
2 School of Geography, Geology and the Environment, Kingston University, Penrhyn Rd, Kingston–Upon–Thames, Surrey, KT1 2EE, UK.
3 East Stroudsburg University, 200 Prospect St., East Stroudsburg, Pennsylvania, 18301–2999, USA.
4 4901 Cherry Tree Bend, Victoria B. C., V8Y 1SI, Canada.
*Correspondence: ecolewis@gmail.com

Abstract - Fish predation can affect amphibian populations. Most examples report invasive fish species and their negative effect on specific amphibians. Here we provide an occurrence of natural fish predation by the tropical gar *Atractosteus tropicus* on the cane toad *Bufo marinus*. We observed physical damage caused by *A. tropicus* attacking *B. marinus*. When adult *B. marinus* were removed from the population subadult toads re-filled the adult's niche, re-supplementing the population. Local observations of predation by fish on amphibians are important to differentiate natural population fluctuations from amphibian population declines.

Key Words - *Bufo marinus*, Predation, Garfish, Population.

Resumen - La depredación de los peces puede afectar a las poblaciones de anfibios. La mayoría de los ejemplos reportan las especies invasoras de peces y su efecto negativo sobre los anfibios. Aquí proveemos la ocurrencia de la depredación natural de los peces *Atractosteus tropicus* en el sapo *Bufo marinus*. Hemos observado los daños físicos causados por *A. tropicus* atacando *B. marinus*. Cuando adultos de *B. marinus* fueron removidos de la población, los sapos subadultos vuelven a ocupar el nicho del adulto, recompletando la población. Observaciones locales de la depredación de los peces en anfibios son importantes para diferenciar las fluctuaciones naturales de la población de la disminución de las poblaciones de anfibios.

Palabras clave - *Bufo marinus*, Predación, Garfish, Población.

*Bufo marinus* (Linnaeus, 1758) (Fig. 1) is Costa Rica's largest amphibian (85–145mm males/90–238mm females SVL), and is found in lowland and premontane zones (Savage 2002). It is one of the world's most studied amphibians due to its successful colonization outside its native range and its negative invasive impacts on foreign fauna (Lever 2001, Phillips and Shine 2005, Urban et al. 2007). Much of the known ecology of *B. marinus* has been described from its natural range in Central and South America (Zug and Zug 1979, Easteal 1986). It is a prolific colonizer of secondary areas and has a home range up to 160m². It is active in moist conditions throughout the year, breeding opportunistically (Strüssmann et al. 1984). Populations of *B. marinus* have been monitored at Caño Palma Biological Station, northeast Costa Rica from 2002-2005 (Lewis 2009).

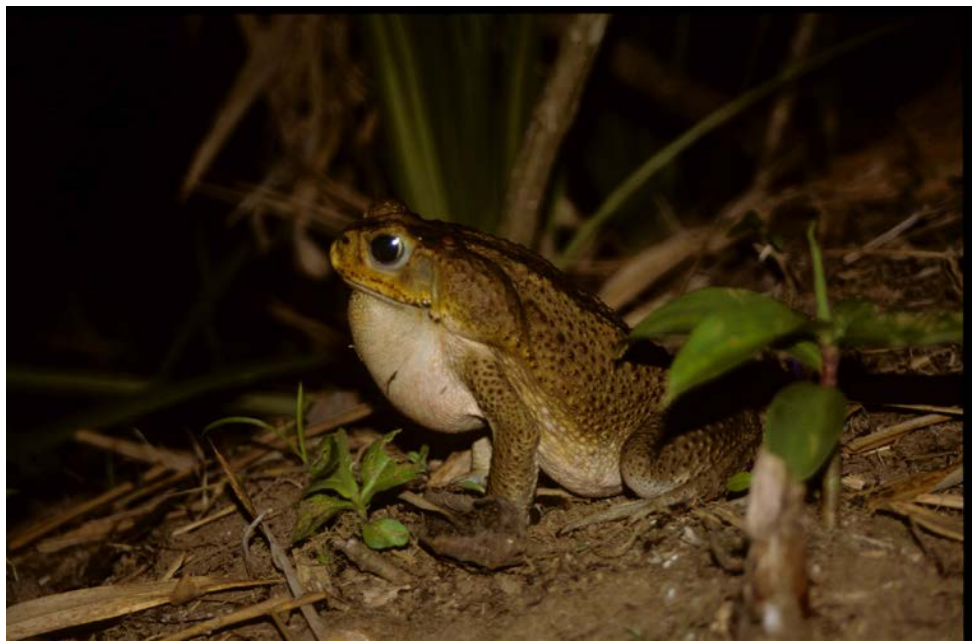

Figure 1 - *Bufo marinus* at Caño Palma Biological Station, Costa Rica.

Caño Palmas forest comprises *Manicaria* swamp forest in a coastal catchment zone on the edge of the Rio Penitencia floodplain boundary (Myers 1990, Lewis 2009, Lewis et al. 2010). Due to its location among the canals of the area, and its high water table, the Caño Palma seasonally floods (Lewis 2009).

Caño Palma Biological Station has a man-made pond in its garden that provides refuge for several predatory fish; *Archocentrus nigrofasciatus* and *Parachromis managuensis* (Cichlidae), *Rhamdia guatemalensis* and *Rhamdia rogersi* (Pimelodidae), *Astyanax aeneus* (Characidae), and *Atractosteus tropicus* (Lepisosteidae). Many of these fish occur in the pond post-flooding annually and remain until the next flood. During their residence these fish have been observed to attack and consume eggs and tadpole of *Rana vaillanti* and *Bufo marinus*. During a flood in November 2004, a young (<60 cm) *Atractosteus tropicus* Gill, 1863 (Fig. 2) became trapped in the station pond. The garfish was 40-50 cm long and likely arrived in the pond post-flood during receding waters. This species uses a bottom-up attack strategy on surface dwelling prey (Bussing 1987). The garfish was observed to prey on small cichlids in the station pond although its survival was questioned due to the pond's limited prey resource. The gar's residence ended upon the arrival of a second flood two weeks later in December 2004.

Five *B. marinus* that were monitored during the population study received injuries during the garfish's residence (Fig. 3a and 3b). Injuries sustained included lacerations to forearm skin and muscular tissue, and complete or partial amputation of the distal end of limbs. All five toads continued to act normally, both feeding, calling, and exhibiting breeding behavior. However, one month after the initial observation of injuries, those toads with limb damage showed signs of weight loss and stress. One specimen died and was found at the pond edge. *A. tropicus* was postulated to be the cause of the injuries. To confirm this, all five amputee toads were collected and deposited in the University of Costa Rica's amphibian collections. Collection and euthanasia was performed using techniques documented in Heyer et al. (1994). Collection was required to discount the possibility of *Ribeiroia* spp. trematodes in the toads, a known parasite of planorbid snails, amphibians and water birds (Johnson and Sutherland 2003). *Ribeiroia* spp. has a cosmopolitan global distribution (Wilson et al. 2005) and induces limb deformity in amphibians when metacercariae become lodged between limb and phalanges joints (Johnson et al. 1999, 2001). After examination of the specimens we determined that the injuries sustained by *B. marinus* were a result of predation and not *Ribeiroia* infection. This was evident from shattered radial and carpal bones under the healed wounds of the toads and no presence of metacercariae in the carpus, metacarpals and phalanges of the specimens.

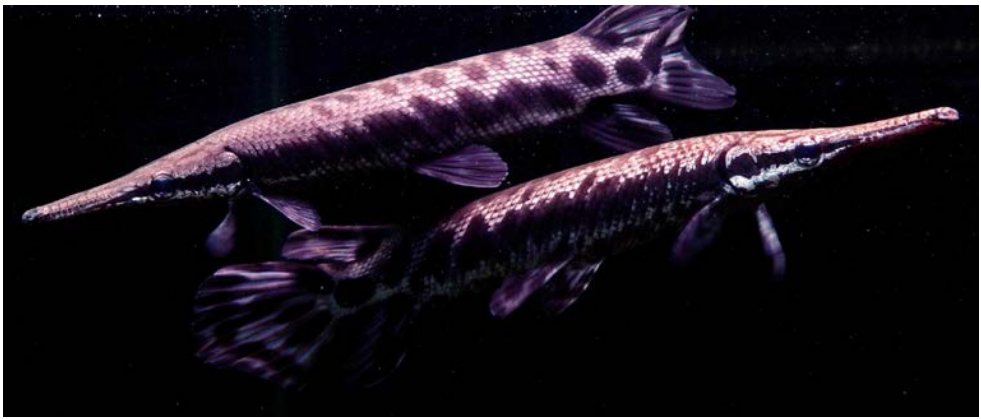

Figure 2 - Young *Atractosteus tropicus*. © Soloman R. David.

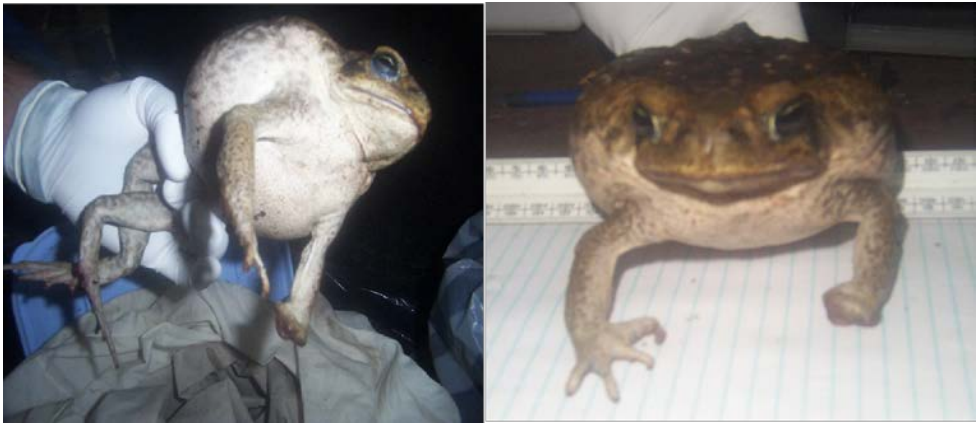

Figure 3a/b - Adult *Bufo marinus* with forelimb injury and healed scar tissue.

An outcome of the removal of the five toads was a subsequent increased recruitment to the pond's toad population. The population of *B. marinus* at the pond had been marked using toe–tipping (Phillott et al. 2007) under permit (MINAE 030846321). The mark-recapture period started in November 2004 and ended in February 2005. The number of toads at the pond varied nightly between three and up to 14. We used an Open Jolly–Seber population estimate to describe how the population fluctuated through time (Fig. 4). The population model was tested and constructed using the program MARK (White and Burnham 1999). Akaike information criterion (AIC) for goodness-of-fit was used to fit the model correctly. Nearly all the toads during November–December 2004 had already been marked. Re-captures were frequent and no unmarked toads had been seen in the population for over two months. From day 1 to day 23 population estimates for all cohorts of *B. marinus* fluctuated between 17 and 93. From days 26 to 31, following the flood and arrival of *A. tropicus* the toad population estimation varied from 0 to 63, signalling a drop in estimated gross population (Fig. 4). Within just two days of collecting the injured toads on day 41, five new adults arrived at the pond and a further nine unmarked subadults were observed on the periphery of the pond. The population estimation continued a decline trend but fluctuated from 0 to 72.

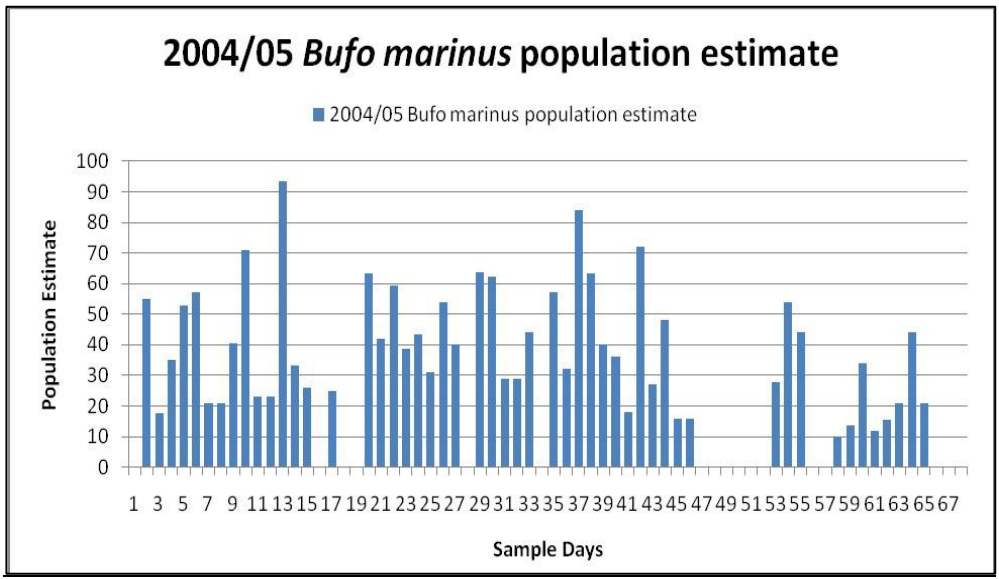

Figure 4 - Open Jolly–Seber population estimates of *Bufo marinus* during 2004/05. Day 39 = collection of adults with injuries.

In tropical floodplain areas like *Manicaria* swamp forest periodic flooding can provide fish greater foraging opportunities (Welcomme 1979, Meffe 1984, Lowe–McConnell 1987, Junk 1997, Drago et al. 2003). During inundation many tropical fish respond to the advantages of increased food supply and obtain higher post-flood growth rates (Talling and Lemoalle 1998). Juveniles and adults of *B. marinus* are preyed upon by select fish, crocodilians, birds, snakes, turtles and mammals that can tolerate Bufotoxin (Chen and Kovarikova 1967, Zug and Zug 1979). Our observations could be the first recorded predation on *B. marinus* by *A. tropicus*. Fish predation on amphibians can influence tropical amphibian community structure and dynamics (Heyer et al. 1975, Sih 1992, Stebbins and Cohen 1995, Kats and Ferrer 2003). Direct effects of predatory fish on amphibians mostly arise from consumption of adults and larvae (Hecnar and M'Closkey 1997, Kiesecker and Blaustein 1998, Smith et al. 1999). Fish presence can also influence the behavior of adult and larvae (Skelly 1992, Anholt et al. 1996), and subsequently affect their abundance and populations (Sih 1992, Brönmark and Edenham 1994, Skelly 1996). Indirect effects between predator and prey can also result in increased competition (Lima and Dill 1990, Morin 1999), change in microhabitat use (Semlitsch and Reyer 1992, Binckley and Re-setarits 2003), altered activity periods (Skelly and Werner 1990, Semlitsch and Reyer 1992), developmental plasticity in larvae (Van Buskirk et al. 1997, Lardner 2000), altered breeding phenology (Kats and Ferrer 2003), induction of early egg-hatching (Warkentin 1995), changes in community dynamics (Fauth 1990, Vredenberg 2004), and synergistic pressures with other aquatic predators, all which increase the likelihood of population decline (Kiesecker and Blaustein 1998). Amphibian responses to fish include chemical cue decisions to avoid water containing fish, developmental strategies and toxicity to make them unpalatable (Kats et al. 1988, Rundio and Olson 2003, Bosch et al. 2006).

Altered behavior by adult *B. marinus* was not observed and adults continued to return to and enter the station pond to breed when *A. tropicus* was resident, thus placing them at risk from predation. The detection of newly recruited individuals was likely to have been the result of the absence of dominant male specimens from the population (the amputees that were collected). Thus recruitment may continue to supply the population in times of stress from temporary predators. *B. marinus* can exist in densities of up to 300/ha although is usually found between 25–30/ha (Zug and Zug 1979). The population estimation at the station pond averaged 39 across the entire period of survey (N = 38.9, AIC 1525.3, Dev. 867.7). The speed of recruitment exhibited by *B. marinus* upon removal of the limb damaged adults may indicate that there is a ready supply of subadults that live on the periphery of the adult territories. These subadults could close the gap on missing adults removed by predation. That the population of *B. marinus* could be affected long-term by presence of predatory fish at the station pond is questionable and invites further study. The effects of predators on small populations of amphibians, at local levels may help further understand their predator–prey relationships in natural environments. Understanding natural fluctuations in amphibian populations, and the interaction of amphibia with other fauna is essential to distinguish between natural population fluctuations and potential declines (Pechman and Wilbur 1997, Alford and Richards 1999, Storfer 2003).

Acknowledgments

We thank the Canadian Organization for Tropical Education and Rainforest Conservation and Xavier Guevara of the Ministerio del Ambiente y Energía Sistema Nacional de Áreas de Conservación, for permits and assistance.